%% file: main.tex
\documentclass[conference]{IEEEtran}
\IEEEoverridecommandlockouts
\usepackage{cite}
\usepackage{amsmath,amssymb,amsfonts}
\usepackage{algorithmic}
\usepackage{graphicx}
\usepackage{textcomp}
\usepackage{xcolor}
\usepackage[utf8]{inputenc}
\usepackage{cite}
\usepackage{times}  
\usepackage{helvet} 
\usepackage{courier}  
\usepackage[hyphens]{url}  
\usepackage{graphicx} 
\usepackage{graphicx}  
\usepackage{amsfonts}
\usepackage{amsmath}
\usepackage{bm} 
\newcommand{\vect}[1]{\boldsymbol{\mathbf{#1}}}
\usepackage{comment}
\usepackage[linesnumbered,ruled]{algorithm2e}

\def\BibTeX{{\rm B\kern-.05em{\sc i\kern-.025em b}\kern-.08em
    T\kern-.1667em\lower.7ex\hbox{E}\kern-.125emX}}
\begin{document}

\title{Constrained Deep Reinforcement Learning for Energy Sustainable Multi-UAV based Random Access IoT Networks with NOMA
}

\author{
    \IEEEauthorblockN{Sami Khairy\IEEEauthorrefmark{1}, Prasanna Balaprakash\IEEEauthorrefmark{2}, Lin X. Cai\IEEEauthorrefmark{1}, Yu Cheng\IEEEauthorrefmark{1}}
    \IEEEauthorblockA{\IEEEauthorrefmark{1}Department of Electrical and Computer Engineering, Illinois Institute of Technology, Chicago, USA }
    \IEEEauthorblockA{\IEEEauthorrefmark{2}Mathematics and Computer Science Division, Argonne National Laboratory, Illinois, USA}
    \IEEEauthorblockA{Email: skhairy@hawk.iit.edu, pbalapra@anl.gov, \{lincai,cheng\}@iit.edu}
}

\maketitle


\begin{abstract}
In this paper, we apply the Non-Orthogonal Multiple Access (NOMA) technique to improve the massive channel access of a wireless IoT network where solar-powered Unmanned Aerial Vehicles (UAVs) relay data from IoT devices to remote servers. Specifically, IoT devices contend for accessing the shared wireless channel using an adaptive $p$-persistent slotted Aloha protocol; and the solar-powered UAVs adopt Successive Interference Cancellation (SIC) to decode multiple received data from IoT devices to improve access efficiency. To enable an energy-sustainable capacity-optimal network, we study the joint problem of dynamic multi-UAV altitude control and multi-cell wireless channel access management of IoT devices as a stochastic control problem with multiple energy constraints. To learn an optimal control policy, we first formulate this problem as a Constrained Markov Decision Process (CMDP), and propose an online model-free Constrained Deep Reinforcement Learning (CDRL) algorithm based on Lagrangian primal-dual policy optimization to solve the CMDP. Extensive simulations demonstrate that our proposed algorithm learns a cooperative policy among UAVs in which the altitude of UAVs and channel access probability of IoT devices are dynamically and jointly controlled to attain the maximal long-term network capacity while maintaining energy sustainability of UAVs. The proposed algorithm outperforms Deep RL based solutions with reward shaping to account for energy costs, and achieves a temporal average system capacity which is $82.4\%$ higher than that of a feasible DRL based solution, and only $6.47\%$ lower compared to that of the energy-constraint-free system. 
\end{abstract}


\section{Introduction}

While internet connectivity plays an increasing role in people's everyday life in densely populated areas, some rural areas and nature fields such as farms, deserts, oceans, and polar regions, typically lack expansive internet coverage. This is because network providers tend to deploy telecommunication infrastructure in areas where providing wireless service is economically profitable. Nevertheless, farmers, environmental agencies, research organizations, defense agencies, and utility companies among many others, have increasing demands for internet connectivity in such under-served areas, to support massive Internet of Things (IoT) based applications ranging from tracking animal health, agricultural growth, and marine life, to surveillance sensors for defense applications and nuclear waste site management, just to name a few. Provisioning wireless internet access for a massive number of IoT devices in under-served areas at cost effective rates is undoubtedly of great interest for governments, businesses, and end customers. 

To support emerging IoT based services, new coverage and distributed channel access solutions should be conceived. Unmanned Aerial Vehicle (UAV) based wireless relays have been proposed to facilitate fast and flexible deployment of communication infrastructure due to their high mobility \cite{lyu2016placement,mozaffari2016efficient,zhao2018deployment,savkin2019method,wu2018joint,xu2018throughput,abd2019deep,liu2019trajectory,hu2018reinforcement}. UAVs equipped with wireless transceivers can fly to a designated area and provision affordable wireless internet connectivity to a massive number of IoT devices, by relaying data to network servers through satellite back-haul links. Moreover, UAVs can dynamically adjust their location in real-time to counter environmental changes and improve system performance.

With the number of IoT devices projected to reach $3.3$ billion by $2021$ \cite{Cisco}, distributed Medium Access Control (MAC) protocols based wireless technologies such as Wi-Fi, Zigbee, and Aloha-based LoRaWAN, are expected to play an important role in provisioning massive IoT access over the unlicensed band in the fifth-generation ($5$G) wireless network era \cite{khairy2018renewal,khairy2019sustainable,el2018lora}. Non Orthogonal Multiple Access (NOMA), which can improve the spectral efficiency by exploiting Successive Interference
Cancellation (SIC) to enable non-orthogonal data
transmissions, is yet another promising solution to enable massive machine type communication (mMTC) in $5$G networks and beyond. Recent works propose to apply power-domain NOMA in slotted-Aloha systems to support mMTC of IoT devices \cite{choi2017noma,seo2018nonorthogonal,choi2018game}. NOMA enabled Aloha-type wireless networks is therefore of significant importance to support massive channel access in UAV-based IoT Networks. 

In this work, we consider solar-powered multi-UAV based wireless IoT networks, where UAVs act as wireless Base Stations (BS) for a massive number of IoT devices. IoT devices contend for access to the shared wireless channel using an adaptive  $p$-persistent slotted Aloha MAC protocol to send data to the UAVs, which relay the received data to the internet backbone through wireless satellite back-haul links. UAVs on the other hand, are equipped with solar cells to replenish the on-board battery, and exploit power-domain SIC to decode multiple users’ transmissions, thus improving the transmission 
efficiency. To enable an energy-sustainable and capacity-optimal massive IoT network, we jointly study dynamic multi-UAV altitude control and NOMA-based multi-cell wireless channel access of IoT devices.
The objective of the stochastic control problem is to maximize the total network throughput of a massive number of IoT devices, characterized by random uplink channel  access and varying wireless channel conditions, and coupled with multiple constraints to ensure energy sustainability of solar-powered UAVs. To the best of our knowledge, our work is the first work to study energy sustainability of a multi-UAV based wireless communication system in support of a massive number of IoT devices with NOMA and random channel access. 

The main contributions of our work can be summarized as follows. First, we formulate the joint problem of multi-UAV altitude control and adaptive random  channel access of massive IoT devices to attain the maximum capacity under energy sustainability constraints of UAVs over a prespecified operating horizon 
as a Constrained Markov Decision Process (CMDP). Second, to learn an optimal control policy for the wireless communication system, we design an online model-free Constrained Deep Reinforcement Learning (CDRL) algorithm based on Lagrangian primal-dual policy optimization to solve the CMDP. A cooperative policy is learned among UAVs which ensures  energy sustainability of UAVs over an operating horizon, while maximizing the total network capacity with NOMA under the probabilistic mutual interference of IoT devices. Third, we compare the performance of our proposed algorithm to two baseline solutions: 1) unconstrained Deep RL (DRL) approach without energy sustainability constraints, and 2) DRL approach with reward shaping to account for energy costs. Our extensive simulations demonstrate that our proposed algorithm yields feasible policies with higher network capacity, and  outperforms baseline solutions which do not guarantee feasible policies. Specifically, the policy learned by our proposed CDRL algorithm achieves a temporal average network capacity that is $82.4\%$ higher than that of a feasible DRL with reward shaping, and $6.47\%$ lower than that of the energy-constraint-free system. Last but not least, we demonstrate that the learned policy, which has been efficiently trained on a small network size, can effectively manage networks with a massive number of IoT devices and varying initial network states. 

The remainder of this paper is organized as follows. A literature survey of related research work and a background of unconstrained and constrained MDPs is given in Section II. The system model is described in Section III. The problem formulation and the proposed CDRL algorithm is presented in Section IV, followed by the simulation setup and performance evaluation results in Section V. Finally our concluding remarks and future work are given in Section VI.

\section{Background and Related Works}
\subsection{UAV based Wireless Networks}
The deployment and resource allocation of UAV-based wireless networks has been studied in many works. In \cite{lyu2016placement}, a polynomial-time algorithm is proposed for successive UAV placement such that the number of UAVs required to provide wireless coverage for a group of ground terminals is minimized and each ground terminal is within the communication range of at least one UAV. The downlink coverage probability for UAVs as a function of the altitude and antenna gain is analyzed in \cite{mozaffari2016efficient}. Based on the circle packing theory, the 3D locations of the UAVs are determined to maximize the total coverage area while ensuring the covered areas of multiple UAVs do not overlap. The work of \cite{zhao2018deployment} studies the problem of multiple UAV deployment for on-demand coverage while maintaining connectivity among UAVs. In \cite{savkin2019method}, a distributed coverage-maximizing algorithm for multi UAV deployment subject to the constraint that UAVs maintain communication is proposed for surveillance and monitoring applications. 

3D trajectory design and resource allocation in UAV based wireless networks have also been studied in \cite{wu2018joint,xu2018throughput,abd2019deep,liu2019trajectory,hu2018reinforcement}. In \cite{wu2018joint}, a mixed integer non-convex optimization problem is formulated to maximize the minimum downlink throughput of ground users by jointly optimizing multi-user communication scheduling, association, UAVs' 3D trajectory, and power control. An iterative algorithm based on block coordinate descent and successive convex optimization techniques is proposed to solve the formulated problem. \cite{xu2018throughput} extends on  \cite{wu2018joint} by considering heterogeneous UAVs so that each UAV can be individually controlled. Machine learning based approaches have also been recently considered for UAV 3D trajectory design. 
In \cite{abd2019deep}, the flight trajectory of the UAV and scheduling of packets are jointly optimized to minimize the sum of 
Age-of-Information (sum-AoI) at the UAV. The problem is modeled as a finite-horizon Markov Decision Process (MDP) with finite state and action spaces, and a DRL algorithm 
is proposed to obtain the optimal policy. \cite{liu2019trajectory} devises a machine learning based approach to predict users’ mobility information, which is considered in the trajectory design of multiple UAVs. A sense-and-send protocol is designed in \cite{hu2018reinforcement} to coordinate multiple UAVs, and a multi-UAV Q-learning based algorithm is proposed for decentralized UAV trajectory design. Scheduling based NOMA systems with a UAV-based BS to serve terrestrial users are considered in \cite{zhao2019joint,nasir2019uav}. In a recent work, the performance of NOMA transmissions in a single-hop random access wireless network is investigated, and an iterative algorithm is proposed to find the optimal transmission probabilities of users to achieve the maximum throughput~\cite{ziru}.

It is worth to mention that all aforementioned works 
consider battery powered UAVs with limited energy storage capacity, which constrains the operating horizon. 
Solar-powered UAVs have great potential to extend the operation time by harvesting solar energy from the sun \cite{morton2015solar,oettershagen2016perpetual}. \cite{lee2017optimal} studies the optimal trajectory of solar-powered UAVs for maximizing the solar energy harvested. In their design, a higher altitude is preferable to maximize harvested energy. On the other hand, \cite{sun2019optimal} studies the trade-off between solar energy harvesting and communication system performance of a single UAV based wireless network. It is shown that in order to maximize the system throughput, the solar-powered UAV climbs to a high altitude to harvest enough solar energy, and then descends to a lower altitude to improve the communication performance. 

The work of \cite{sun2019optimal} considers downlink wireless resource allocation in a centralized scheduling-based and interference-free wireless network with a single UAV. Deploying one solar-powered UAV may lead to a communication outage when the UAV ascends to high altitudes to replenish its on-board battery. On the other hand, scheduling-based networks usually suffer from the curse of dimensionality and do not well scale to massive IoT networks, as signaling overheads scale up with the network size. As such, there is a growing interest in wireless networks with NOMA and  decentralized random access MAC protocols, such as Aloha-type MACs adopted in LoRaWAN networks \cite{el2018lora}. Analyzing NOMA performance and modeling the probabilistic channel interference caused by uplink transmissions in multi-cell random channel access wireless networks is a very challenging task as it is mathematically intractable. Machine learning provides a data driven approach for system design, and can be used to investigate these challenging wireless systems. Thus motivated, in this work we study energy sustainability of solar-powered multi-UAV based massive IoT networks with random-access and NOMA. We design an online model-free CDRL algorithm for dynamic control of UAVs' altitude and random wireless channel access management. By deploying multiple UAVs, we demonstrate that is possible to learn a cooperative policy in which multiple UAVs take turns to charge their battery and provision uninterrupted wireless service.

\subsection{Constrained Deep Reinforcement Learning}
One of the primary challenges faced in reinforcement learning is the design of a proper reward function which can efficiently guide the learning process. Many real world problems are multi-objective problems in which conflicting objectives should be optimized. A common approach to handling multi-objective problems in RL is to combine the objectives using a set of coefficients \cite{mannor2004geometric}. With this approach, there exist a set of optimal solutions for each set of coefficients, known as the Pareto optimal solutions \cite{van2014multi}. In practice, finding the set of coefficients which leads to the desired solutions is not a trivial task. For many problems, it is more natural to specify a single objective and a set of constraints. The CMDP framework is the standard formulation for RL problems involving constraints \cite{altman1999constrained}. Optimal policies for CMDPs can be obtained by solving an equivalent linear programming formulation \cite{altman1999constrained}, or via multi time-scale dynamic-programming based algorithms  \cite{bhatnagar2012online,bhatnagar2010actor,borkar2005actor,geibel2012learning,geibel2005risk}. Such methods may not be applicable to large scale problems or problems with continuous state-action space. Leveraging recent advances in deep learning and policy search methods \cite{schulman2015trust}, some works devise multi-time scale algorithms for solving RL problems in presence of constraints \cite{tessler2018reward,liang2018accelerated,fu2018risk,chow2017risk, achiam2017constrained}. Broadly speaking, these methods are either based on Lagrangian relaxation \cite{tessler2018reward,liang2018accelerated,fu2018risk,chow2017risk} or constrained policy optimization \cite{achiam2017constrained}. In Lagrangian relaxtion based method, primal and dual variables are updated at different time-scales using gradient ascent/descent. In these methods, constraint satisfaction is guaranteed at convergence. On the other hand, in \cite{achiam2017constrained} an algorithm is proposed where constraint satisfaction is enforced in every step throughout training. Our proposed algorithm is based on the Proximal Policy Optimization (PPO) \cite{schulman2017proximal}, which is a highly stable state-of-the-art on-policy model-free RL algorithm that adopts the Lagrangian relaxation based approach to handle multiple constraints. To the best of our knowledge, our work is the first to demonstrate successful policy learning in environments with multiple constraints, and policy transferability among wireless networks of different scales. 

\subsection{Background}
In this subsection, unconstrained and constrained MDPs, which are the classical formalization of sequential decision making and define the interaction between a learning agent and its environment in RL and constrained RL, are introduced.  
\subsubsection{Markov Decision Process}
An infinite horizon MDP 
with discounted-returns is defined as a tuple $(\mathcal{S},\mathcal{A},\mathcal{P}, \mathcal{P}_0,\mathcal{R},\zeta)$,  where $\mathcal{S}$ and $\mathcal{A}$ are finite sets of states and actions, respectively, $\mathcal{P}:\mathcal{S}\times \mathcal{A} \times \mathcal{S} \rightarrow [0,1]$ is  the model's state-action-state transition probabilities, and $\mathcal{P}_0:\mathcal{S} \rightarrow [0,1]$ is the initial distribution over the states, $\mathcal{R}: \mathcal{S} \times \mathcal{A} \rightarrow \mathbb{R}$, is the immediate reward function which guides the agent through the learning process, and $\zeta$ is a discount factor to bound the cumulative rewards and trade-off how far or short sighted the agent is in its decision making. Denote the transition probability from state $s_n=i$ at time step $n$ to state $s_{n+1}=j$ if action $a_n=a$ is chosen by $P_{ij}(a) := P(s_{n+1} = j | s_n = i, a_n = a)$ 
The transition probability from state $i$ to state $j$ is therefore, $p_{ij} = P(s_{n+1} = j | s_n = i) = \sum_a P_{ij}(a) \pi(a|i)$, where $\pi(a|i)$ is a stochastic policy which maps states to actions. The state-value function of state $i$ under policy $\pi$ is long-term expected discounted returns starting in state $i$ and following policy $\pi$ thereafter, 
\begin{equation} \label{stateValue}
\resizebox{1\hsize}{!}{$
V_\pi(i) = \sum_{n=1}^\infty  \sum_{j, a} \zeta^{n-1} P^\pi(s_n = j, a_n = a | s_0 = i) \mathcal{R}(j,a), \forall i \in \mathcal{S}$}
\end{equation}
Denote the initial distribution over the states by the vector $\vect{\beta}$, where $\beta(i) = P(s_0=i), \forall i \in \mathcal{S}$. The solution of an MDP is a Markov stationary policy $\pi^*$ that maximizes the inner product $\langle\, \vect{V}_\pi, \vect{\beta} \rangle$,

\begin{equation}\label{eq:sys1}
\underset{\pi}{\text{max}} ~~~~~~ \sum_{n=1}^\infty \sum_{j, a} \zeta^{n-1} P^\pi(s_n = j, a_n = a) \mathcal{R}(j, a)
\end{equation}
There are several approaches to solve \eqref{eq:sys1}, including dynamic programming based methods such as value iteration and policy iteration,  \cite{sutton2018reinforcement}, in addition to linear programming based methods \cite{kallenberg2011markov}. When the model's dynamics, i.e., transition probabilities, are unknown, the Reinforcement Learning (RL) framework can be adopted to find the optimal policies. It is worth to mention that when the agents learns the optimal state-value function and/or the policy as parameterized Deep Neural Networks (DNNs), the agent is commonly referred to as a Deep RL (DRL) agent. There exists a significant body of works with state-of-the-art algorithms to solve the RL problem, which vary by design from value-based methods \cite{mnih2013playing}, policy-based methods \cite{sutton2000policy,schulman2015trust,schulman2017proximal}, to hybrid actor-critic type algorithms \cite{silver2014deterministic,lillicrap2015continuous,fujimoto2018addressing,mnih2016asynchronous,haarnoja2018soft}. 

\subsubsection{Constrained Markov Decision Process}
In constrained MDPs (CMDPs), additional immediate cost functions $\mathcal{C}_k: \mathcal{S} \times  \mathcal{A} \rightarrow \mathbb{R}$ are augmented, such that a CMDP is defined by the tuple $(\mathcal{S},\mathcal{A},\mathcal{P}, \mathcal{P}_0,\mathcal{R},\mathcal{C}, \zeta)$ \cite{altman1999constrained}.  The state-value function is defined as in unconstrained MDPs \eqref{stateValue}. In addition, the infinite-horizon discounted-cost of a state $i$ under policy $\pi$ is defined as,
\begin{equation}
\resizebox{1\hsize}{!}{$
C^k_\pi(i) = \sum_{n=1}^\infty \sum_{j, a} \zeta^{n-1}  P^\pi(s_n = j, a_n = a | s_o = i) \mathcal{C}_k(j, a), \forall i \in \mathcal{S}, \forall k $.}
\end{equation}
The solution of a CMDP is a markov stationary policy $\pi^*$ which maximizes $\langle\, \vect{V}_\pi, \vect{\beta} \rangle$  subject to the constraints $\langle\, \vect{C}_\pi^k, \vect{\beta} \rangle \leq E_k, \forall k$, 
\begin{subequations}\label{eq:system}
\begin{align}
\underset{\pi}{\text{max}} ~~~~~~ &\sum_{n=1}^\infty \sum_{j, a} \zeta^{n-1}  P^\pi(s_n = j, a_n = a) \mathcal{R}(j, a) \tag{\ref{eq:system}}\\
\label{eq:systemA}
& \sum_{n=1}^\infty  \sum_{j, a} \zeta^{n-1} P^\pi(s_n = j, a_n = a) \mathcal{C}_k(j, a) \leq E_k,~\forall k
\end{align}
\end{subequations}

Solving for feasible and optimal policies in CMDPs is more challenging compared to unconstrained MDPs, and requires extra mathematical efforts. CMDPs can be solved by defining an appropriate occupation measure and constructing a linear program over this measure, or alternatively by using a Lagrangian relaxation technique in which the CMDP is converted into an equivalent unconstrained problem,
\begin{equation}
\begin{aligned}
\underset{\pi}{\text{max}} ~ &\underset{\vect{\eta} \geq 0}{\text{min}} ~ \mathcal{L}(\pi,\vect{\eta}) \\ &= \underset{\pi}{\text{max}} ~ \underset{\vect{\eta}\geq 0}{\text{min}}~
    \langle \vect{V}_\pi, \vect{\beta} \rangle - \sum_k \eta_k \big(\langle \vect{C}_\pi^k, \vect{\beta} \rangle - E_k \big)
\end{aligned}
\end{equation}
and invoking the minimax theorem,
\begin{equation} \label{uncons:lag}
    \underset{\pi}{\text{max}} ~ \underset{\vect{\eta} \geq 0}{\text{min}} ~ \mathcal{L}(\pi,\vect{\eta}) =  \underset{\vect{\eta} \geq 0}{\text{min}} ~ \underset{\pi}{\text{max}} ~ \mathcal{L}(\pi,\vect{\eta})
\end{equation}
The right hand side of \eqref{uncons:lag} can be solved on two-time scales: on a faster time scale gradient-ascent is performed on state-values to find the optimal policy for a given set of Lagrangian variables, and on a slower time scale, gradient-descent is performed on the dual variables \cite{altman1999constrained}. Past works explore this primal-dual optimization approach for CMDPs with known model dynamics and tabular-based RL methods with unknown model dynamics \cite{bhatnagar2012online,bhatnagar2010actor,borkar2005actor,geibel2012learning,geibel2005risk}. In the realm of deep RL where policies and value functions are parameterized neural networks, recent works which apply primal-dual optimization for generic benchmark problems are emerging \cite{tessler2018reward,liang2018accelerated,fu2018risk,chow2017risk,achiam2017constrained}. None of these works, however, apply primal-dual optimization techniques in the wireless networking domain. Practical wireless networking systems admit multiple constraints, which might be conflicting. This incurs extra difficulty for policy search and optimization. Applying constrained RL for wireless networking problems is therefore a challenging issue that needs to be investigated.


\section{System Model} \label{sec:sysmod}
\begin{figure}
    \centering
    \includegraphics[width=0.85\linewidth]{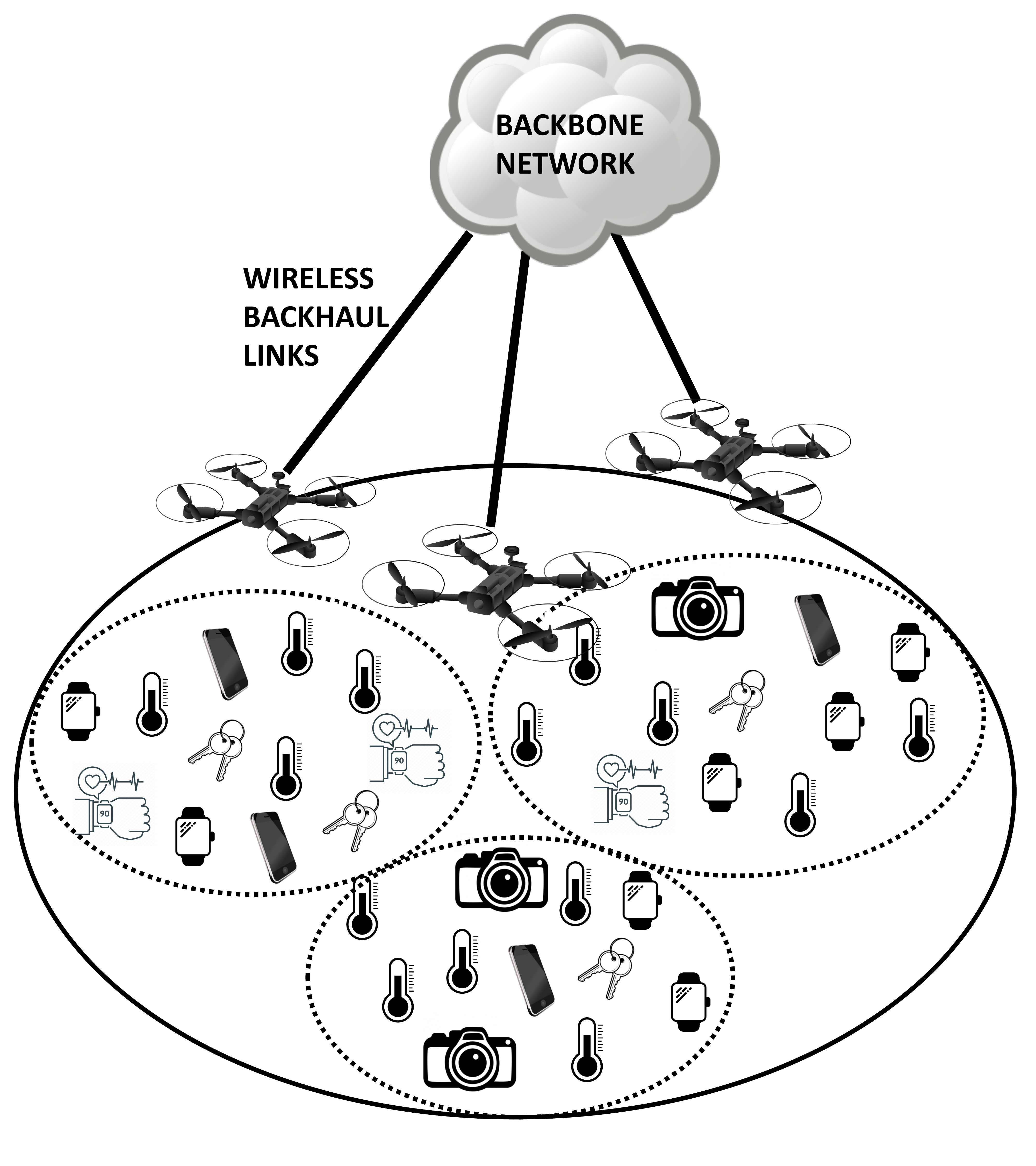}  \\  \scriptsize{(a) Network Model} \\ 
    \includegraphics[width=.98\linewidth]{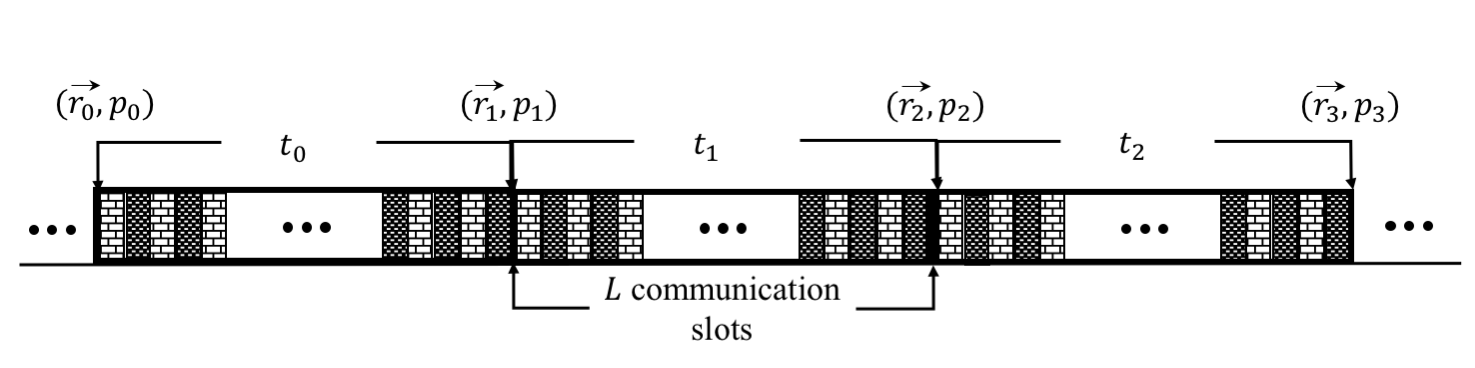} \\ \scriptsize{(b) Time Structure} \\
    \caption{ Multi-cell UAV based  wireless IoT network} \label{fig:sysmod}
  \end{figure}

Consider a multi-UAV based IoT network consisting of $M$ UAVs and $N$ IoT devices, where the UAVs collect data from a massive deployment of IoT devices, as shown in Figure \ref{fig:sysmod}(a). Let $\mathcal{M}=\{1,\cdots, M\}$ be the set of UAVs, and $\mathcal{N}=\{1,\cdots, N\}$ be the set of IoT devices. UAVs are connected via wireless back-haul links to a central controller, which controls the altitude of each UAV and manages the access parameters of wireless IoT devices. IoT devices are independently and uniformly distributed (i.u.d.) across a deployment area $\mathbb{A}$. Let the locations of IoT devices be $\{\hat{x}^i,\hat{y}^i\}_{i=1}^N$. Each IoT device is served by the closest UAV. Denote the subset of IoT devices which are associated with UAV $m$ by $\mathcal{N}_m \subset \mathcal{\mathcal{N}}$, $|\mathcal{N}_m|\leq N$, $\bigcup_{m=1}^M \mathcal{N}_m = \mathcal{N}$, $\mathcal{N}_i \cap \mathcal{N}_j= \phi, \forall i\neq j \in \mathcal{M}$. Time is slotted into fixed-length discrete time units indexed by $n$. For instance, the $n$-th time slot is $[t_n, t_{n+1})$, where $t_{n+1} - t_n = \Delta t,~\forall n$. Each time slot $n$ is further divided into $L$ communication sub-slots of length $\frac{\Delta t}{L}$ each, as shown in \ref{fig:sysmod}(b). Denote the $l$-th communication sub-slot in slot $n$ by $t_{n}^l$, $l=\{0, \cdots, L-1\}$. During these communication sub-slots, IoT devices contend for channel access based on an adaptive $p$-persistent slotted Aloha MAC protocol. In this protocol, an IoT device waits until the
beginning of a communication sub-slot before attempting to access the channel with probability $p$, which is adapted every time slot by the central controller based on network dynamics. IoT devices transmit uplink  data to their associated UAV with a fixed transmission power of $P_{TX}$ watts, and are traffic-saturated, i.e., there is always a data packet ready for transmission. 

Denote the location of UAV $m$ during time slot $n$ by $r_m(t_n) = (x^m, y^m, z^m_n) $. In our system model, $(x^m, y^m), \forall m \in \mathcal{M}$, is first determined based on the IoT device distribution on the ground using the popular Lloyd's K-means clustering algorithm with runtime $\mathcal{O}(2NM)$ \cite{lloyd1982least}\footnote{Lloyd's K-means clustering algorithm is an iterative algorithm to determine a set of $K$ centroids given a large set of IoT device locations $\{ \hat{x}^i,\hat{y}^i \}_{i=1}^N$, so as to minimize the within-cluster variance (sum of squared distances to cluster centroid), $\text{min}_{\{(x^m,y^m)\}} \sum_{m=1}^M \sum_{i\in \mathcal{N}_m} ||(\hat{x}^i,\hat{y}^i)-(x^m,y^m)||^2$. 
The impacts of adopting other schemes for determining $(x^m, y^m), \forall m \in \mathcal{M}$ on network performance will also be investigated.}.
Let $u^{i,m}_n=\{0,1\}$ indicate whether IoT device $i$ associates with UAV $m$ during time slot $n$. If UAV $m$ located at $r_m(t_n)$ during the $n$-th time slot is the closest to IoT device $i$, $u^{i,m}_n=1$, and $u^{i,l}_n=0, \forall l \neq m$. The power of the signal transmitted by a wireless IoT device $i$ to a UAV $m$ is subject to independent Rayleigh channel fading,  $h_{i,m}(t_n)$\footnote{The statistical channel state information, $h_{i,m}(t_n)$, is assumed to be quasi-static and is fixed during a time slot $n$.}, and a distance-dependent free-space path-loss $d_{i,m}^{-\alpha}(t_n)$, where $\alpha$ is the path-loss exponent, and $d_{i,m}^{-\alpha}(t_n)$ is the propagation distance between IoT device $i$ and UAV $m$, $d_{i,m}(t_n) = \sqrt{ (x^m-\hat{x}^i)^2 + (y^m-\hat{y}^i)^2 + (z^m_n)^2 }$. The received power at UAV $m$ from IoT device $i$ in a communication sub-slot $t_n^l$ is,
\begin{equation} \label{rxPwr}
\begin{gathered}
\begin{aligned}
P^{i,m}_{RX}(t_n^l) = 
\begin{cases}
\mathcal{\hat{I}}_i(t_n^l) c_0 h_{i,m}(t_n) P_{TX}d_0^{-\alpha},~d_{i,m}(t_n)\leq d_0, \\
\mathcal{\hat{I}}_i(t_n^l) c_0 h_{i,m}(t_n) P_{TX} d_{i,m}^{-\alpha}(t_n), d_{i,m}(t_n)\geq d_0,
\end{cases}
\end{aligned}
\end{gathered}
\end{equation} 
where  $c_0 = \frac{\lambda^\alpha}{(4\pi)^\alpha}$ is a constant which depends on the wavelength of the transmitted signal, $\lambda$, and $d_0$ is a reference distance. $\mathcal{\hat{I}}_i(t_n^l)$ is a Bernoulli random variable with parameter $p(t_n) \in (0,1]$ which indicates whether IoT device $i$ transmits during communication sub-slot $t_{n}^l$, i.e., $\mathcal{\hat{I}}_i(t_n^l)=1$ with probability $p(t_n)$. The UAV $m$ first decodes the signal with the highest signal power under the interference from all other IoT devices involved in the NOMA transmissions. Without loss of generality, IoT devices $\mathcal{I}_m(t_n^l) = \{i | \mathcal{\hat{I}}_i(t_n^l)=1 \}$ are sorted in the descending order of their received signal strength at UAV $m$, such that $i=1$ is the IoT device with the highest received signal to interference plus noise (SNIR) at UAV $m$, and $i=2$ is the IoT device with the second highest received SNIR at UAV $m$ \footnote{We consider the two highest received signals to trade-off NOMA gain and SIC decoding complexity for uplink transmissions.}. The highest received SNIR at UAV $m$ in a communication sub-slot $t_{n}^l$ is therefore,  
\begin{equation}
    \text{SNIR}_{1,m}(t_n^l) = \frac{P^{1,m}_{RX}(t_n^l)}{n_0 + \sum_{k \in \mathcal{I}_m(t_n^l)\setminus 1} P^{k,m}_{RX}(t_n^l)},
\end{equation}
where $n_0$ is the noise floor power. Similarly, the second highest received SNIR at UAV $m$ in a communication sub-slot $t_{n}^l$ is,
\begin{equation}
    \text{SNIR}_{2,m}(t_n^l) = \frac{P^{2,m}_{RX}(t_n^l)}{n_0 + \sum_{k \in \mathcal{I}_m(t_n^l)\setminus \{1,2\} } P^{k,m}_{RX}(t_n^l)},
\end{equation}
UAV $m$ can decode the signal with $\text{SNIR}_{1,m}(t_n^l)$ if 
\begin{enumerate}
    \item user 1 is associated with UAV $m$ during communication sub-slot $n$, $u_n^{1,m}=1$, and,
    \item $\text{SNIR}_{1,m}(t_n^l)$ is larger than the SNIR threshold, i.e.,  $U(\text{SNIR}_{1,m}(t_n^l))=\text{SNIR}_{1,m}(t_n^l)$,
\end{enumerate}
where $U(.)$ is a thresholding function to maintain a minimum quality of service, 
\begin{equation} \label{snrthres}
\begin{gathered}
\begin{aligned}
U(\text{SNIR}_{i,m}(t_n^l))=
\begin{cases}
0,~\text{SNIR}_{i,m}(t_n^l) < \text{SNIR}_\text{Th}, \\
\text{SNIR}_{i,m}(t_n^l), \text{SNIR}_{i,m}(t_n^l) \geq \text{SNIR}_\text{Th}.
\end{cases}
\end{aligned}
\end{gathered}
\end{equation} 
In addition, UAV $m$ can decode the signal with $\text{SNIR}_{2,m}(t_n^l)$ if
\begin{enumerate}
    \item $\text{SNIR}_{1,m}(t_n^l)$ is successfully decoded,
    \item user 2 is associated with UAV $m$ during communication sub-slot $n$, $u_n^{2,m}=1$, and,
    \item $\text{SNIR}_{2,m}(t_n^l)$ is larger than the SNIR threshold, i.e., $U(\text{SNIR}_{2,m}(t_n^l))=\text{SNIR}_{2,m}(t_n^l)$
\end{enumerate}
The sum rates of the received data at UAV $m$ in communication sub-slot $t_n^l$ is,
\begin{equation}
\begin{aligned}
    G_{m}(t_n^l) =& \mathcal{W}\text{log}_2\Big(1+U(\text{SNIR}_{1,m}(t_n^l))u_n^{1,m}\Big) +\\
    &\mathcal{W}\text{log}_2\Big(1+U(\text{SNIR}_{2,m}(t_n^l))u_n^{1,m}u_n^{2,m}e_n^{1,m}\Big) 
\end{aligned}
\end{equation}
where $\mathcal{W}$ is the transmission bandwidth, and $e_n^{1,m} = 1$ if  $U(\text{SNIR}_{1,m}(t_n^l))=\text{SNIR}_{1,m}(t_n^l)$ and $0$ otherwise. The total network capacity in any given system slot $t_n$,
\begin{equation} \label{Capacity}
    \mathbb{G}(t_n) =  \sum_{l=0}^{L-1} \sum_{m \in \mathcal{M}} G_{m}(t_n^l)
\end{equation}

UAVs are equipped with solar panels, which harvest solar energy to replenish the on-board battery. The  attenuation  of  solar  light passing through a cloud can be modeled based on \cite{sun2019optimal}, 
\begin{equation}
    \phi(d^{cloud}) = e^{-\beta_c d^{cloud}}
\end{equation}
where $\beta_c \geq 0$ denotes the absorption coefficient of the cloud, and $d^{cloud}$ is the distance that the solar light travels through the cloud. Following \cite{sun2019optimal} and the references therein, the solar energy harvested by UAV $m$ during time slot $n$ can be modeled as, 
\begin{equation} \label{HarvE}  
\resizebox{1\hsize}{!}{
    $E_{\text{H}}^m(t_n) = \begin{cases}
    \psi \Tilde{S}\Tilde{G} \Delta t,~~~~~~~~~~~~~~~~~~~~~~~~\frac{z_n^m+z^m_{n+1}}{2} \geq z_{high} \\
    \psi \Tilde{S}\Tilde{G} \phi(z_{high}-\frac{z_n^m+z^m_{n+1}}{2})\Delta t, ~~~~z_{low} \leq \frac{z^m_n+z^m_{n+1}}{2} < z_{high} \\
    \psi \Tilde{S}\Tilde{G} \phi(z_{high}-z_{low})\Delta t,~\frac{z^m_n+z^M_{n+1}}{2} < z_{low}
    \end{cases}$}
\end{equation}
where $\psi$ is a constant representing the energy harvesting efficiency, $\Tilde{S}$ is  the area of solar panels, and $\Tilde{G}$ denotes the average solar radiation intensity on earth. $z_{high}$ and $z_{low}$ are the altitudes of upper and lower boundaries of the cloud. During time-slot $n$, UAV $m$ can cruise upwards or downwards from $r_m(t_n)$ to $r_m(t_{n+1})$. The energy consumed by UAV $m$ during time slot $n$ \cite{sun2019optimal} is,  

\begin{equation} \label{ConsE}
\begin{aligned}
E_{\text{C}}^m(t_n) =& \left( \frac{W^2/(\sqrt{2}\rho A)}{4^{0.25}V_z} + Wv_z +  P_{\text{static}} + P_{\text{antenna}}\right) \Delta t, \\
& v_z = \frac{z_{n+1}^m - z_n^m}{\Delta t}
\end{aligned}
\end{equation}
where, $V_z = \sqrt{\frac{W}{2 \rho A}}$, $W$ is the weight of the UAV, $\rho$ is air density, and $A$ is the total area of UAV rotor disks. $P_{\text{static}}$ is the power consumed for maintaining the operation of UAV, and $P_{\text{antenna}}$ is the power consumed by the receiving antenna. It is worth to mention that cruising upwards consumes more power than cruising downward and hovering.

Denote the battery energy storage of UAV $m$ at the beginning of slot $n$ by $B_m(t_n)$. The battery energy in the next slot is given by, 
\begin{equation}
\resizebox{0.9\columnwidth}{!}{$
    B_m(t_{n+1}) = \text{min} \{ \left[B_m(t_{n})+  E_{\text{H}}(t_n) - E_{\text{C}}^m(t_n)+\mathbb{B}(t_n)\right]^+, B_{\text{max}} \}$},
\end{equation}
where $\mathbb{B}(t_n),\forall n$, are independent zero-mean gaussian random variables with variance $\sigma^2_B$ which characterizes the randomness in the battery evolution process, and $\left[~ \right]^+$ denotes the positive part.

\section{Problem Formulation and Proposed CDRL Algorithm}
In this work, we investigate discrete-time UAV altitude control and random channel access management for a multi-UAV based IoT network with NOMA. In order to maximize the total network capacity under stochastic mutual interference of IoT devices while ensuring the energy sustainability of UAVs over the operating horizon $H$, the central controller decides on the altitude of each UAV $m,~\forall m \in \mathcal{M}$, at the beginning of each slot $n$, $z_n^m$, as well as the channel access probability $p(t_n)$ of IoT devices considering the potential access gain provisioned by NOMA. The channel access probability will be broadcast to IoT devices through beacons at the beginning of each time slot, and IoT devices then adapt their random channel access parameter accordingly. 

The problem of maximizing the total network capacity while ensuring energy sustainability of each UAV is a constrained stochastic optimization problem over the operating horizon due to the stochastic channel model and random channel access in the multi-cell IoT network. Offline solutions are generally impractical because it is hard to mathematically track probabilistic mutual interference caused by the uplink transmissions of IoT devices with random access, stochastic channel conditions, and SIC decoding at the UAVs. Hence, we first formulate this problem as a CMDP, and design an online Constrained Deep Reinforcement Learning (CDRL) algorithm, to find an energy sustainable capacity-optimal policy to control the altitude of each UAV and the channel access probability of IoT devices.

\subsection{CMDP Formulation} \label{sec:cmdp_form}

To enable continuous control of UAVs altitudes and channel access probability, we consider parametrized DNN based policies with parameters $\vect{\theta}$ and state-value function with parameters $\vect{\Theta}$ henceforth. In this subsection, we formulate the joint problem of UAVs altitude control and random channel access of IoT devices as a discrete-time CMDP with continuous state and action spaces as follows, 
\begin{enumerate}
\item  $\forall s_n \in \mathcal{S}$, 
\begin{equation} \nonumber
 \begin{aligned}
 s_n = \bigcap_\mathcal{M} &\Bigg\{z_n^m,\cdots, z_{n-h_k}^m,  B_m(t_n),\cdots,B_m(t_{n-h_k}), \\
 &P\big(\text{SNIR}_{1,m}(t_n^l)\geq \text{SNIR}_{\text{Th}}\big), \\ &P\big(\text{SNIR}_{2,m}(t_n^l)\geq \text{SNIR}_{\text{Th}}\big), \\ &\mathbb{E}\big[\text{SNIR}_{1,m}(t_n^l)|\text{SNIR}_{1,m}(t_n^l)\geq \text{SNIR}_{\text{Th}}\big],\\ &\mathbb{E}\big[\text{SNIR}_{2,m}(t_n^l)|\text{SNIR}_{2,m}(t_n^l)\geq \text{SNIR}_{\text{Th}}\big],\\
&\text{Var}\big[\text{SNIR}_{1,m}(t_n^l)|\text{SNIR}_{1,m}(t_n^l)\geq \text{SNIR}_{\text{Th}}\big], \\ &\text{Var}\big[\text{SNIR}_{2,m}(t_n^l)|\text{SNIR}_{2,m}(t_n^l)\geq \text{SNIR}_{\text{Th}}\big]\Bigg\},\\
 \end{aligned} 
\end{equation}
 i.e., the state space encompasses $\forall m$, the current altitude of UAV $m$ along with $h_k$ historical altitudes, current battery energy of UAV $m$ along with $h_k$ historical battery energies, probability the highest and second highest received SNIRs from associated users at UAV $m$ is greater than or equal to $\text{SNIR}_\text{Th}$, the mean of 
 the highest and second highest received SNIRs from associated users at UAV $m$ given that they are greater than or equal to the SNIR threshold, and the variance of 
 the highest and second highest received SNIRs from associated users at UAV $m$ given that they are greater than or equal to the SNIR threshold. Here the mean and variance are calculated over the $L$ communication sub-slots. 
 
\item $\forall a_n \in \mathcal{A}$, $a_n = \bigcap_\mathcal{M}\{\Delta z_n^m\} \cap \{p(t_{n+1}) \}$, where $\Delta z_n^m= z_{n+1}^m - z_n^m$, i.e., the action space encompasses the altitude displacement of each UAV between any two consecutive time  slots, and the random channel access probability in the next system slot.

\item $\mathcal{R}(s_n, a_n) = \frac{\mathbb{G}(t_n)}{H}$, i.e., the immediate reward at the end of each time slot $n$ is the total network capacity during slot $n$, normalized by the operating horizon $H$. 

\item $\mathcal{C}_m(s_n, a_n) = \frac{B_m(t_n) - B_m(t_{n+1})}{B_{max}}$, $\forall m$, i.e.,  the immediate cost at the end of each slot $n$ is the change in the battery energy between any two consecutive time slots, which is caused by the displacement of each UAV $m$, normalized by the maximum battery energy.

\item $E_m = -B_{min}, \forall m$, i.e., the upper bound on the long-term expected cost is the negative of the minimum desired battery energy increase at the end of the planning horizon over the initial battery energy. 
\end{enumerate}

Based on this formulation, the objective is to find a Markov policy 
$\pi_{\vect{\theta}_\pi}$ which maximizes the long-term expected discounted total network capacity, while ensuring energy sustainability of each UAV $m$ over an operating horizon, 
\begin{equation} \label{eq:CRL}
    \begin{aligned}
        \underset{\vect{\theta}_\pi}{\text{max}}~~~& \mathbb{E}_{\pi_{\vect{\theta}}}^\beta \Big[\sum_{n=0}^\infty \zeta^n \mathbb{G}(t_n)\Big] \\
        & \mathbb{E}_{\pi_{\vect{\theta}}}^\beta \Big[\sum_{n=0}^H B_m(t_n) - B_m(t_{n+1}) \Big] \leq -B_{min}, \forall m
    \end{aligned}
\end{equation}
\eqref{eq:CRL} exhibits trade-offs between total system capacity and energy sustainability of UAVs. For instance, a UAV hovering at a higher altitude above the cloud cover can harvest more solar energy to replenish its on-board battery storage, as given by \eqref{HarvE}. However, at higher altitudes, the received signal power at a UAV from IoT devices will be smaller due to the log-distance path loss model, and consequently, the system capacity will be smaller. The converse is true, that is, when a UAV hovers at lower altitudes, network capacity is improved, yet solar energy harvesting is heavily attenuated. In addition, based on the network topology or the location of the UAVs at any time slot $n$, a certain spatial gain and NOMA overload can be achieved. An optimal stochastic control policy for altitude control of UAVs and channel access management of IoT devices should be therefore learned online. In the following subsection, we propose an online CDRL algorithm to solve \eqref{eq:CRL}.

\subsection{Proposed CDRL Algorithm}

To solve \eqref{eq:CRL} in absence of the  state-action-state transition probabilities of the Markov model, we adopt the RL framework in which an autonomous agent learns an optimal policy by repeated interactions with the wireless environment \cite{sutton2018reinforcement}. The wireless environment provides the agent with rewards and costs signals, which the agents exploit to further improve its policy.
\begin{figure}
    \centering
    \includegraphics[width=1\linewidth]{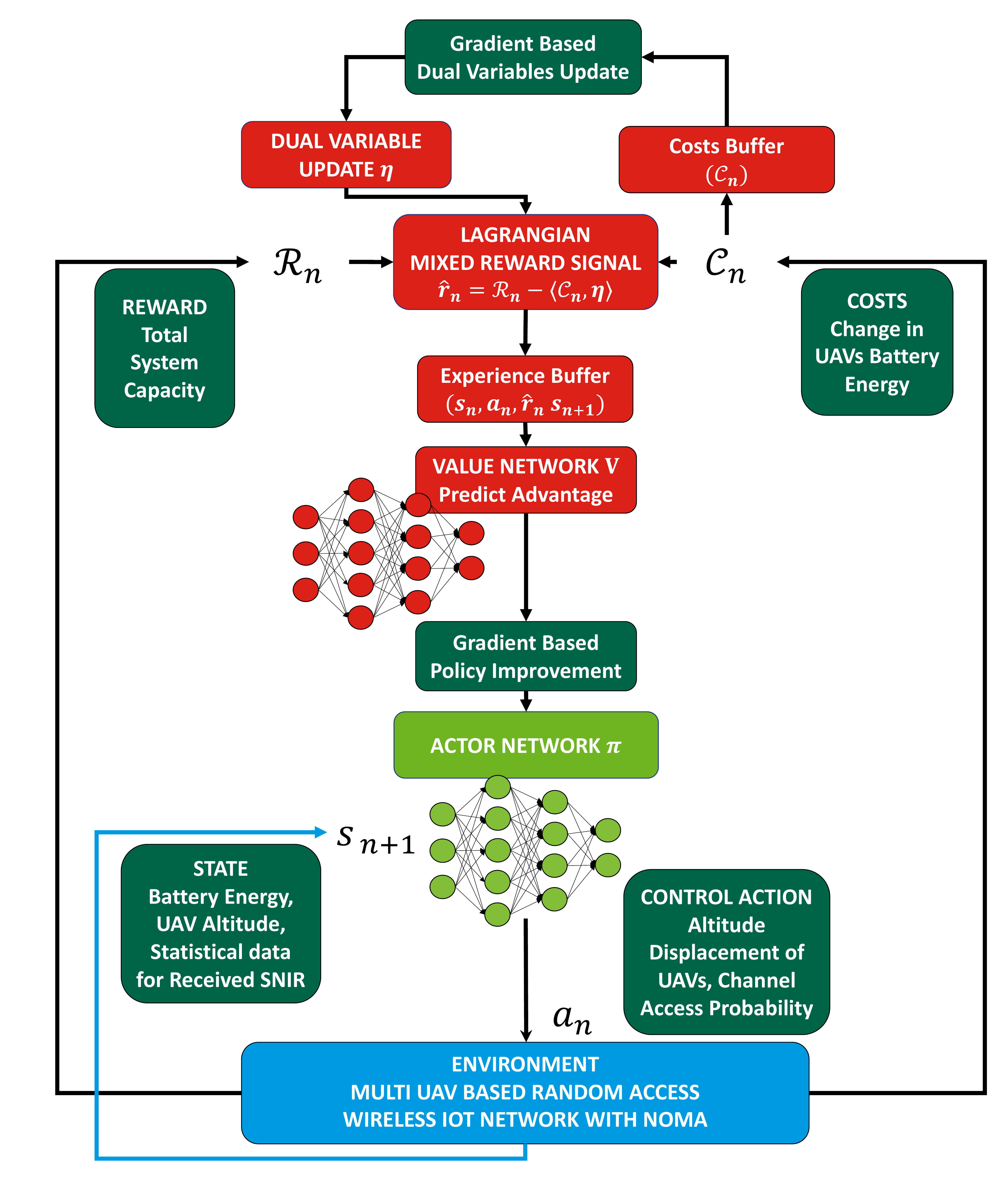}
    \caption{Proposed constrained deep reinforcement learning architecture} \label{fig:arch}
 \end{figure}
Our proposed algorithm is based on the state-of-the-art Proximal Policy Optimization (PPO) algorithm \cite{schulman2017proximal}, and leverages the technique of primal-dual optimization \cite{liang2018accelerated}. The architecture of our proposed algorithm is shown in Figure \ref{fig:arch}. 

In the proposed CDRL algorithm, parameterized DNN of the policy $\pi_{\vect{\theta}}(a|s)$ is learned by maximizing the PPO-clip objective function, which is a specially designed clipped surrogate advantage objective that ensures constructive policy updates,
\begin{equation} \label{ppo}
\resizebox{1\hsize}{!}{$
O^\text{clip}(\vect{\theta}) = \hat{\mathbb{E}}_n\Big[\text{min}( \frac{\pi_{\vect{\theta}} (a_n|s_n)}{\pi_{\vect{\theta}_\text{old}}(a_n|s_n)} \hat{A}_n, \text{clip}(\frac{\pi_{\vect{\theta}} (a_n|s_n)}{\pi_{\vect{\theta}_\text{old}}(a_n|s_n)}, 1+\epsilon, 1-\epsilon)\hat{A}_n) \Big]$},
\end{equation}
where $\vect{\theta}$ are the policy neural network parameters, $\epsilon$ is a clip fraction, and $\hat{A}_n$ is the generalized advantage estimator (GAE) \cite{schulman2015high}\footnote{The advantage function is defined as the difference between the state-action value function and the value function, $A(s_n,a_n)=Q(s_n,a_n)-V(s_n)$. GAE makes a compromise between bias and variance in estimating the advantage.}, 
\begin{equation} \label{gae}
    \hat{A}_n = \sum_{l=0}^\infty (\zeta \xi)^l \big(\hat{\mathcal{R}}_{n+l} + \zeta V_{\vect{\Theta}}(s_{n+l+1}) - V_{\vect{\Theta}}(s_{n+l}) \big).
\end{equation}
Clipping in \eqref{ppo} acts as a regularizer which controls how much the new policy can go away from the old one while still improving the training objective. In order to further ensure reasonable policy updates, we adopt a simple early stopping method in which gradient optimization on \eqref{ppo} is terminated when the mean KL-divergence between the new and old policy reaches a predefined threshold $\text{KL}_{\text{Th}}$. In \eqref{gae}, $\hat{\mathcal{R}}_{n}=\hat{\mathcal{R}}_{n}(s_n,a_n,\vect{\eta})$ is a Lagrangian penalized reward signal \cite{tessler2018reward}, 
\begin{equation} \label{pen:rew}
    \hat{\mathcal{R}}(s_n,a_n,\vect{\eta}) = \mathcal{R}(s_n,a_n) - \sum_m \eta_m \mathcal{C}_m(s_n,a_n).
\end{equation}
In our proposed algorithm, the Lagrangian penalty multipliers are updated adaptively according to policy feasibility by performing gradient descent on the original constraints. Towards this end, we define the following loss function which is minimized with respect to $\vect{\eta}$,
\begin{equation} \label{lag:loss}
\resizebox{1\hsize}{!}{$
    O^{P}(\vect{\eta})= \sum_m \eta_m \text{clip}\Big(-B_{min} - \mathbb{E}_{\pi_{\vect{\theta}}}^\beta \Big[\sum_{n=0}^H B_m(t_n) - B_m(t_{n+1}) \Big], -\infty, 0 \Big)$}
\end{equation}
Finally, the state-value function is learned by minimizing the mean squared error loss against the policy's discounted rewards-to-go,
\begin{equation} \label{mse:V}
    O^\text{V}(\vect{\Theta}) = \hat{\mathbb{E}}_n\Big[ \Big(V_{\vect{\Theta}}(s_n) - \sum_{l=0}^\infty \zeta^l \hat{\mathcal{R}}_{n+l}(s_{n+l},a_{n+l}) \Big)^2 \Big],
\end{equation}
The optimization in our proposed algorithm is performed over three time-scales, on the fastest time scale, the state-value function is updated by minimizing \eqref{mse:V}, then the policy is updated by maximizing \eqref{ppo} on the intermediate time-scale, 
and finally, the Lagrangian multipliers are updated on the slowest time-scale by minimizing \eqref{lag:loss}. Optimization time-scales are controlled by choosing the maximum learning rate of the stochastic gradient optimizer used, e.g., ADAM \cite{kingma2014adam}, as well as the number of gradient steps performed at the end of each training epoch. The full algorithmic procedure for training the CDRL agent is outlined in Algorithm $1$.

\begin{algorithm} \label{Alg1}
 \SetKwInOut{Input}{Input}
    \LinesNotNumbered
    \Input{Initial policy network parameters $\vect{\theta}$, initial value network parameters $\vect{\Theta}$, initial Lagrange multipliers $\vect{\eta}=\vect{0}$  }
\For{$\text{epoch}=0,1, \cdots$ }
{
\For{$n=0,1, \cdots$, $H$}
{
Observe initial state $s_n$ \\
Sample action $a_n \sim \pi_{\vect{\theta}}(a_n|s_n)$\\
Take action $a_n$\\
Receive reward $\mathcal{R}(s_n,a_n)$, $M$ costs $\mathcal{C}_m(s_n,a_n)$, and new state $s_{n+1}$\\
Compute penalized reward $\hat{\mathcal{R}}(s_n,a_n,\vect{\eta})$ using \eqref{pen:rew} \\
Store transition $(s_n,a_n,\hat{\mathcal{R}}(s_n,a_n,\vect{\eta}),s_{n+1})$ in policy training buffer\\
Store $\mathcal{C}_{k,\forall k}(s_n,a_n)$ in Lagrange multiplier training buffer
}
Compute rewards to go\\
Compute advantage estimate $\hat{A}_n$ using GAE \eqref{gae} and current value network\\
\For{$k=0,1,\cdots$}
{
Update policy parameters ${\vect{\theta}}_k$ via stochastic gradient ascent with ADAM on the PPO-Clip objective \eqref{ppo} \\
Compute KL-divergence between new policy and old policy\\
Break if KL-divergence hits $\text{KL}_{\text{Th}}$\\
}
\For{$k=0,1,\cdots$}
{
Fit the value network via stochastic gradient descent with ADAM on \eqref{mse:V}
}
Update Lagrangian multipliers via stochastic gradient descent with ADAM on  \eqref{lag:loss}
}
    \caption{Constrained PPO-Clip}
\end{algorithm}

The adopted policy is a parameterized stochastic Gaussian policy, 
\begin{equation}
    \pi_{\vect{\theta}} (a_n|s_n) = \frac{1}{\sigma(s_n,\vect{\theta_\sigma})\sqrt{2\pi}} \text{exp} \Big(- \frac{(a_n - \mu(s_n,\vect{\theta_\mu}))^2}{2\sigma(s_n,\vect{\theta_\sigma})^2} \Big).
\end{equation}
where $\vect{\theta_\mu}$ are the DNN parameters for the mean of the policy, and $\vect{\theta_\sigma}$ are the DNN parameters for the variance of the policy. At the beginning of training, the variance of the policy network encourages exploration. As training progresses, the variance of the policy is reduced due to maximizing \eqref{ppo} and the policy shifts slowly towards a deterministic policy.

\textbf{\textit{CDRL Implementation and Training}}: A fully connected multi-layer perceptron network with three hidden layers for both the policy and value networks are used. Each hidden layer has $128$ neurons. $Tanh$ activation units are used in all neurons. The range of output neurons responsible for the altitude displacement of each UAV is linearly scaled to $[\Delta z_{min},\Delta z_{max}]$ in order to limit the maximum cruising velocity, while the range of the output neuron in charge of the random channel access probability is linearly scaled to $[0,\frac{2}{N}]$. The weights of the policy network are heuristically initialized to generate a feasible policy. The variance of the Gaussian policy is state-independent, $\sigma(s_n,\vect{\theta_\sigma)} =\vect{\theta_\sigma}$, with initial value $\vect{\theta^0_\sigma} = e^{-0.5}$. Training has been performed over $1000$ epochs, where each epoch corresponds to $32$ episodes, and each episode corresponds to trajectories of length $H$ time steps. At the end of each episode, the trajectory is cut-off and the wireless system is reinitialized. Episodes in each epoch are rolled-out in parallel by $32$ Message Passing Interface (MPI) ranks. 
After each MPI rank completes its episodic roll-out, Lagrangian primal-dual policy optimization is performed locally as outlined in Algorithm $1$, based on the averaged gradients of the MPI ranks, such that the DNN parameters $\vect{\theta}$, $\vect{\Theta}$, and the Lagrange multipliers $\vect{\eta}$, are synchronized among the $32$ MPI ranks during training. At the end of training, the trained policy network corresponding to the mean of the learned Gaussian policy, $\mu(s_n, \vect{\theta}_\mu)$, is used to test its performance through the simulated environment.

\section{Performance Evaluation}

We have developed a simulator in Python for the solar-powered multi-UAV based Wireless IoT network with NOMA described in section \ref{sec:sysmod}, and implemented the proposed CDRL algorithm based on OpenAI's implementation of PPO ~\cite{openaippo}. We trained the proposed CDRL agent in a multi-cell wireless IoT network of $M=2$ solar-powered UAVs and $N=200$ IoT devices. IoT devices were deployed independently and uniformly within a grid of $[0,0] \times [1500,500]$m. The two UAVs were initially deployed at $(250,250,750)$m and $(750,250,1250)$m with $50\%$ initial battery energy, i.e., $111$ Wh. Although we have investigated the impacts of different initial UAV deployments on network performance, note that the $x$ and $y$ coordinates of the two UAVs, i.e., $(250,250)$m and $(750,250)$m, are minimizers of the sum of squared planar distances between IoT devices and the closest UAV, as determined by Lloyd's K-means clustering algorithm for the uniform random deployment of IoT devices on the ground. UAVs were allowed to cruise vertically between $500$m and $1500$m. As a baseline for comparison, we have compared the performance of the proposed CDRL algorithm with unconstrained PPO agent without energy sustainability constraints, and unconstrained PPO agents that accounts for energy costs via fixed reward shaping (RLWS), where the reward signal was $\mathcal{\hat{R}}(s_n,a_n,\vect{\eta}) = \mathcal{R}(s_n,a_n)-\eta_1\mathcal{C}_1(s_n,a_n)-\eta_2\mathcal{C}_2(s_n,a_n)$, and $\eta_1=10,\eta_2=10$  or $\eta_1=0,\eta_2=10$. The main simulation parameters for the experiments  are outlined in Table \ref{table:simparam}. 

\begin{table}[ht]  
\caption{Simulation Parameters}   
\centering                          
\begin{tabular}{c c | c c }            
\hline\hline                        
Parameter  & Value &  Parameter & Value \\ [0.5ex] 
\hline                              
$P_{TX}$ & $30~dBm$  & $B_{min}$ & $22$ Wh    \\
$\alpha$  & $2$  & $\psi$ & $0.4$    \\
$\Delta t$ & $10s$    & $\Tilde{S}$ & $1m^2$   \\
$f_0$ & $900$ MHz  &$\Tilde{G}$ & $1367 W/m^2$   \\
$d_0$ & $1m$   &$W$ &  $39.2 kg*m/s^2$   \\
$h_{i,m}$ & $exp(1)$ & $\rho$ &  $1.225kg/m^3$ \\
$n_0$ & $-80dBm$ &  $A$ &  $0.18m^2$\\
$\mathcal{W}$ & $1Hz$  & $P_{static}+P_{antenna}$ &  $5$ watts \\
$\text{SNIR}_\text{Th}$ & $10dB$  &  $\epsilon$ & $0.2$ \\
$L$ & $1000$ & $h_k$ & $5$ \\
$\beta_c$ & $0.01$ & $H$ & $360$ (1hr) \\
$z_{high},z_{low}$ & $1.3,0.7$km & $\Delta z_{min},\Delta z_{max}$ & $-40m,40m$  \\
$z_{min},z_{max}$ & $0.5,1.5$km &  $\xi$ & $0.97$    \\
$B_{max}$ & $222$ Wh & $\vect{\eta}_\text{lr}$   & $3\times 10^{-3}$  \\
$\zeta$ & $0.999$  & $\vect{\Theta}_\text{lr}$ &  $10^{-3}$ \\
$\text{KL}_\text{Th}$ & $0.01$  & $\vect{\theta}_\text{lr}$  &  $3\times 10^{-4}$ \\
$\sigma_B^2$ & $500$  &  & \\
\hline                              
\end{tabular}          \label{table:simparam}   
\end{table} 

\begin{figure}
    \centering
    \includegraphics[width=0.95\linewidth]{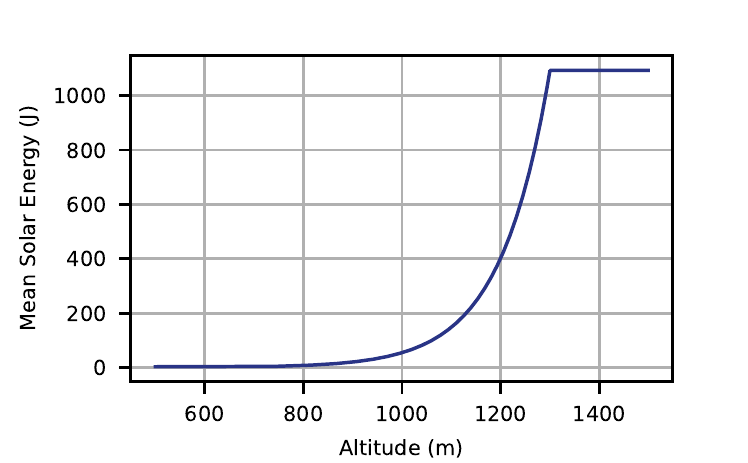} 
    \caption{Solar energy output of solar panels versus UAV altitude. Solar energy decays exponentially through the cloud cover between$1300$m and $700$m.} \label{fig:0}
  \end{figure}

The output energy of the solar panels as a function of the altitude according to \eqref{HarvE} is shown in Figure \ref{fig:0}. It can be seen that above the cloud cover at $1300$m, the output solar energy reaches the highest, and it attenuates exponentially through the cloud until $700$m is reached. Below $700$m, the output solar energy is zero. UAVs cruising above $1300$m harvest the most solar energy, while UAVs cruising below $700$m drain their battery energy the fastest. Note that hovering at low altitudes reduces distance-dependant path-loss and improves the wireless communication rates, however, it is not energy sustainable for UAVs. 

\begin{figure}[h]
    \centering
    \includegraphics[width=0.95\linewidth]{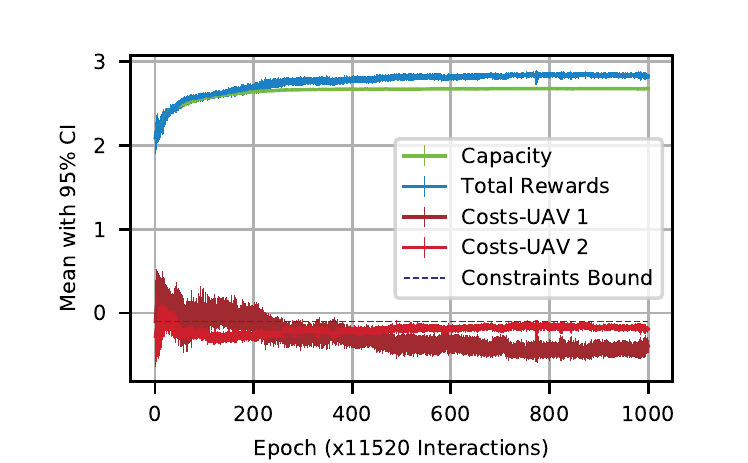}  \\  \scriptsize{(a) Learning Curves} \\ 
    \includegraphics[width=.95\linewidth]{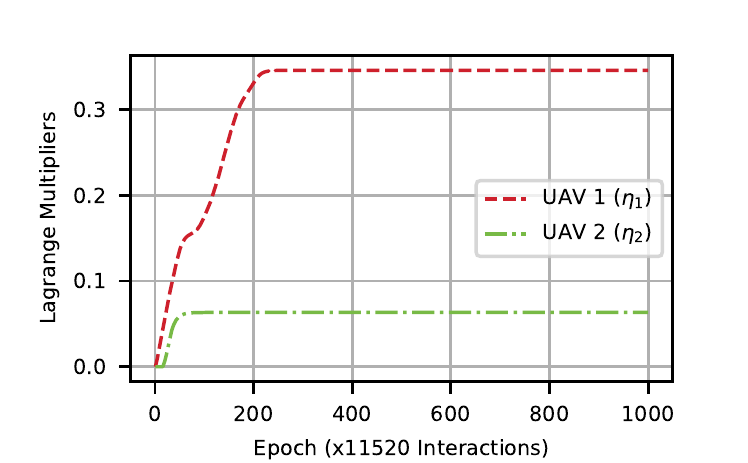} \\ \scriptsize{(b) Lagrange Multipliers} \\
    \caption{Training results of the proposed CDRL agent} \label{fig:1}
  \end{figure}

The learning curves of the trained CDRL agent  are shown in Figure \ref{fig:1}. It can be seen in Figure \ref{fig:1}(a) that the CDRL agent becomes more experienced as training progresses, collecting higher expected total rewards. In addition, the CDRL agent becomes more experienced in satisfying energy constraints, learning policies whose expected costs fall bellow $-B_{min}/B_{max}=-0.1$, which means that the learned policy results in energy increase in the battery of each UAV by at least $B_{min}$ at the end of the flight horizon. On the other hand, the convergence of the Lagrangian multipliers to non-negative values during the training of the proposed CDRL algorithm  is shown in Figure \ref{fig:1}(b). It can be observed that the two cost constraints are penalized differently, which is primarily due to the different initial conditions and different terminal states. 
The Lagrangian multiplier of the energy constraint corresponding to UAV 1 is larger than that of UAV 2, which means that the energy constraint of UAV 1 is satisfied farther from the constraint boundary compared to that of UAV 2. Based on this observation, it is therefore expected that UAV 1 will end up its flight with more harvested energy in its battery compared to UAV 2. 

\begin{figure}
    \centering
    \includegraphics[width=.95\linewidth]{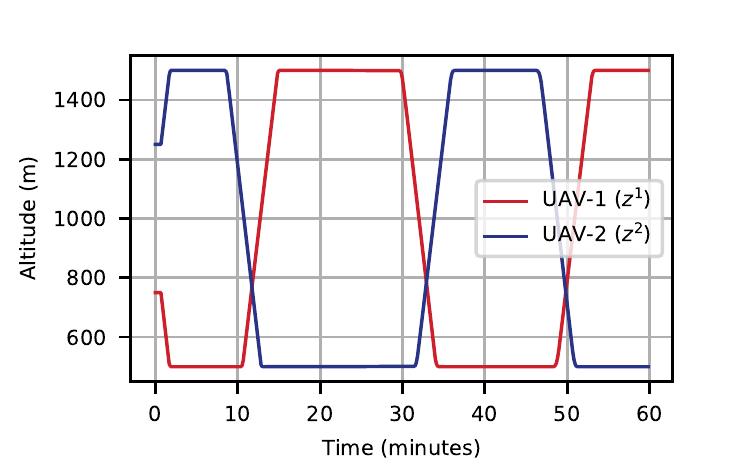} \\ \scriptsize{(a) UAV Altitudes vs Time} \\
    \includegraphics[width=.95\linewidth]{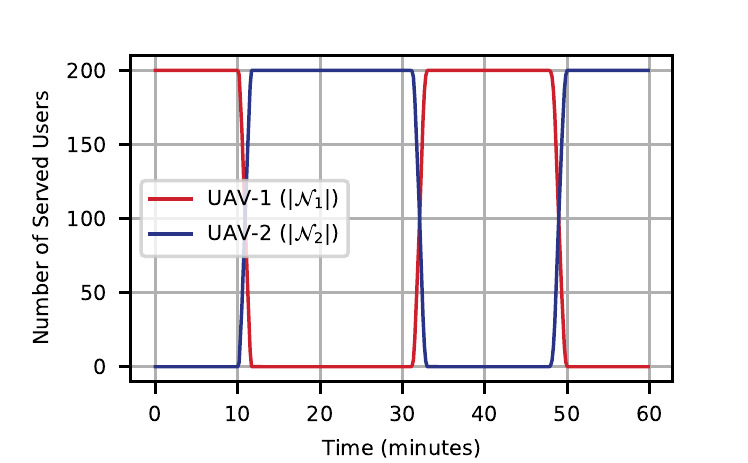} \\ \scriptsize{(b) UAV IoT Device Association vs Time} \\
    \includegraphics[width=.95\linewidth]{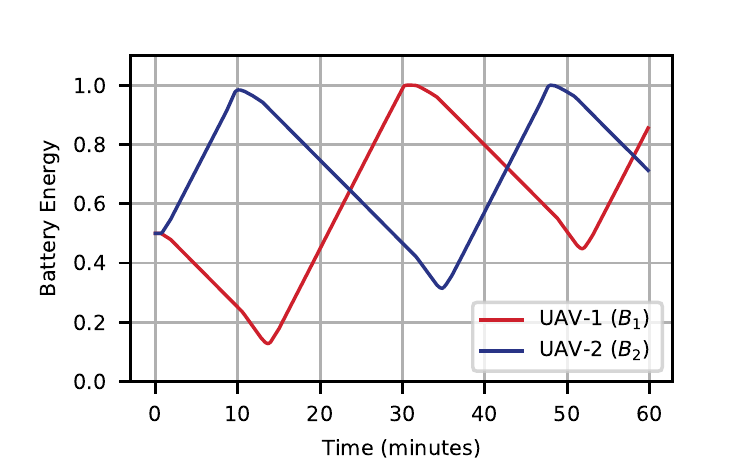} \\ \scriptsize{(c) UAV Battery Energy vs Time} \\    
    \includegraphics[width=0.95\linewidth]{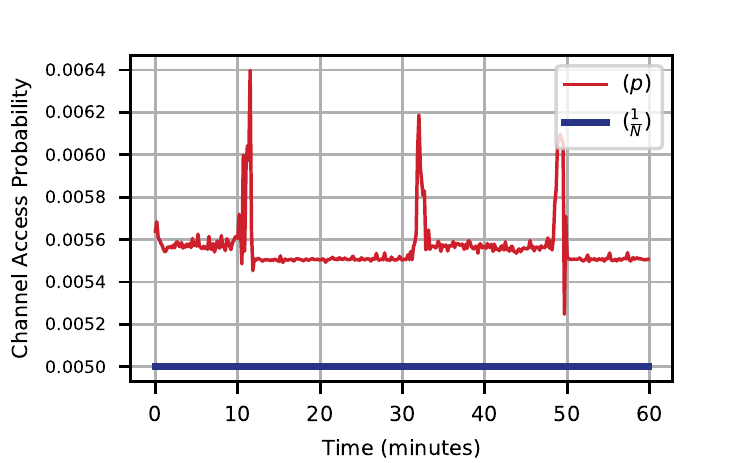}  \\  \scriptsize{(d) Channel Access Probability vs Time} \\ 
    \caption{Learned policy by the CDRL agent during the operating horizon of the two UAVs} \label{fig:3}
  \end{figure}

\begin{figure}
    \centering
    \includegraphics[width=0.95\linewidth]{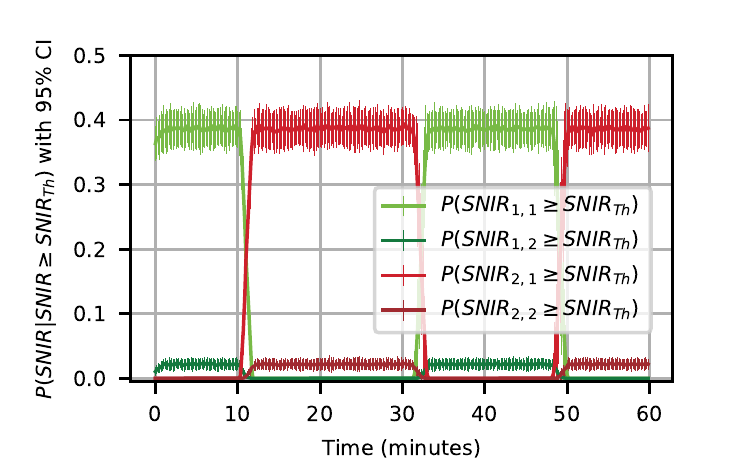}  \\  \scriptsize{(a) Probability SNIR is equal or larger than $\text{SNIR}_{Th}$ } \\ 
    \includegraphics[width=.95\linewidth]{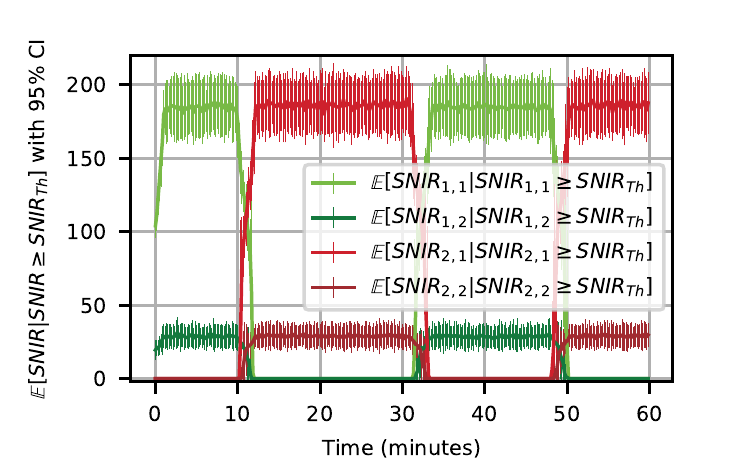} \\ \scriptsize{(b) Conditional Expected SNIR} \\
    \caption{NOMA performance based on the learned CDRL policy during the operating horizon of the two UAVs} \label{fig:4}
  \end{figure}

The learned policy by our proposed CDRL algorithm is shown in Figure \ref{fig:3}. It can be seen from Figures \ref{fig:3}(a) and \ref{fig:3}(b) that the CDRL agent learns a policy in which the two UAVs take turns in cruising upwards to recharge their on-board batteries, and in serving IoT devices deployed on the ground. UAV $2$ first climbs up to recharge its battery, while UAV $1$ descends down to improve communication performance for IoT devices on the ground. Since IoT devices are associated with the closest UAV, in this case, all IoT devices are associated with UAV $1$. When UAV $2$'s battery is fully charged, it descends down gradually to switch roles with UAV $1$: UAV $2$ becomes the BS with which all IoT devices are associated, while UAV $1$ climbs up to recharge its battery. Such a policy ensures that the battery energy of the two UAVs is not drained throughout the operating horizon as can be seen from Figure \ref{fig:3}(c). Note that the terminal energy in UAV's 1 battery is higher than that of UAV 2, which is expected based on the observation that the Lagrange multiplier corresponding to UAV's 1 energy constraint is larger than that of UAV 2. In Figure \ref{fig:3}(d), the random channel access probability based on the learned CDRL policy is shown. It can be observed that when either of the two UAVs is fully serving all the $200$ IoT devices, the wireless networking system is overloaded with $p>1/N$, thanks to NOMA. Note that $p=1/N$ is the optimal transmission probability in single-cell $p$-persistent slotted Aloha systems without NOMA. The channel access probability is dynamically adapted when the two UAVs cruise upward and downward to exchange roles in the wireless system. At times when both UAVs have associated users, the channel access probability can be seen to spike higher to maintain NOMA overload, as can be observed from Figures \ref{fig:4}(a)-(b). It can be seen from these two figures that NOMA's gain is higher in steady states when all IoT devices are associated with the same UAV, compared to transient states when the two UAVs exchange roles and are both serving IoT devices. This is because it is less likely that the second highest received SINR to a UAV is from within the same cell at times when both UAVs provision wireless service. By deploying multiple UAVs, it is therefore possible to learn a cooperative policy in which UAVs take turns to charge their battery and provision uninterrupted wireless service as Figure \ref{fig:4}(b) shows.

\begin{figure}
    \centering
    \includegraphics[width=0.95\linewidth]{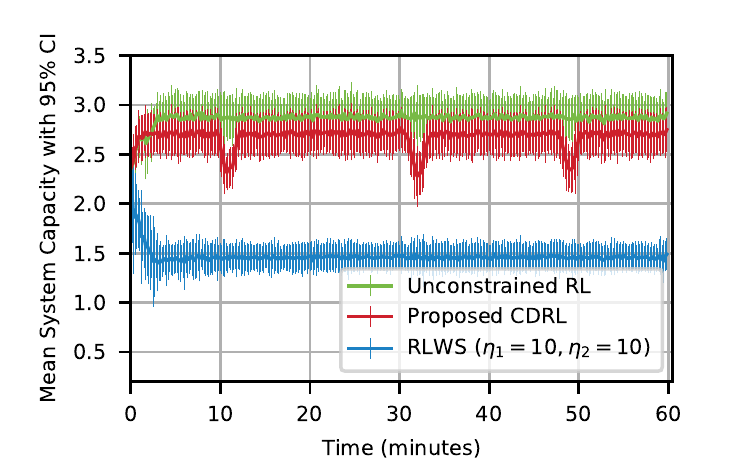}  \\  \scriptsize{(a) System Capacity vs Time} \\ 
    \includegraphics[width=.95\linewidth]{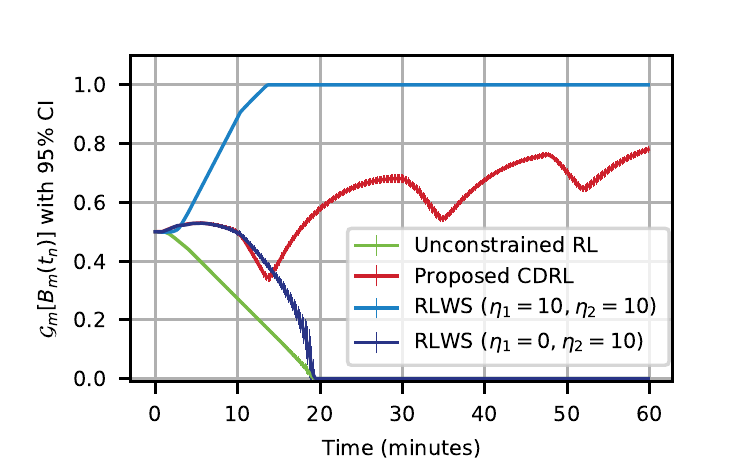} \\ \scriptsize{(b) Sum Battery Energy vs Time} \\
    \caption{Performance comparison of the proposed CDRL algorithm with DRL based solutions with reward shaping} \label{fig:2}
  \end{figure}

\begin{figure}
    \centering
    \includegraphics[width=0.95\linewidth]{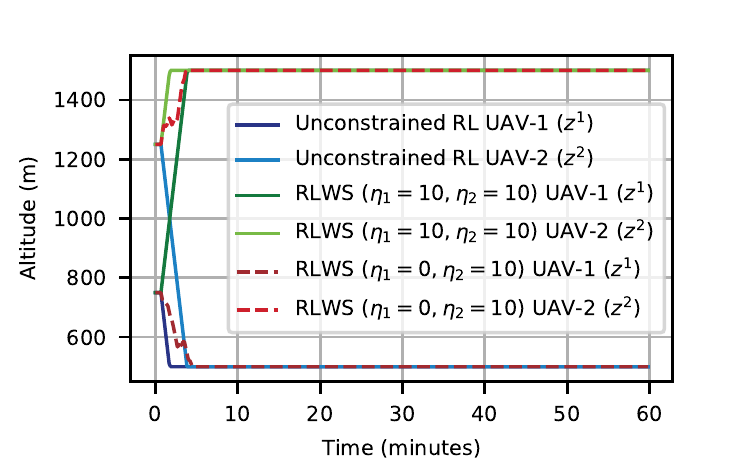} 
    \caption{Learned altitude policy by baseline algorithms} \label{fig:5}
  \end{figure}

The performance comparison of our proposed CDRL algorithm with two other baseline schemes in terms of the achieved mean total network capacity with $95\%$ confidence interval versus flight time is shown in Figure \ref{fig:2}(a). The statistical results are based on $32$ roll-outs of the learned policy in the simulated wireless IoT environment with NOMA. It can be seen that the proposed CDRL agent learns a policy whose achieved temporal average network capacity is $82.4\%$ higher compared to the policy learned by the conservative RLWS agent with $\eta_1=\eta_2=10$. Compared to the RL agent without energy constraints, only $6.47\%$ of the achievable temporal average system capacity is sacrificed in order to maintain energy sustainability of UAVs. In Figure \ref{fig:2}(b), the geometric mean of the battery energy of the two UAVs, defined as $\mathcal{G}_m[B_m(t_n)]= \big( \prod_{m=1}^M B_m(t_n) \big)^{\frac{1}{M}}$, is plotted versus flight time. Note that the geometric mean is chosen as a single measure to characterize battery energy of the UAVs. If the battery of any UAV is exhausted, $B_m(t_n)=0$, the geometric mean will be $\mathcal{G}_m[B_m(t_n)]=0$. It can be seen that the policy learned by the RLWS agent with $\eta_1=\eta_2=10$ is the most conservative, fully recharging the batteries of both UAVs by the $13$-th minute. On the other hand, the unconstrained agent learns a policy which is indifferent to battery energy, leading to energy depletion of at least one UAV by the $20$-th minute. This is reflected in Figure \ref{fig:5}, where the altitude trajectories based on the policies learned by the unconstrained RL agent and the RLWS agents are shown. The policy of the unconstrained agent descends the two UAVs to the lowest allowable altitude of $500$m. This policy is communication-performance bound, achieving the highest network capacity as shown in Figure \ref{fig:2}(a). On the other hand, the policy of the RLWS agent with $\eta_1=\eta_2=10$ attempts to minimize the energy costs and maintain a high battery energy at all times. Hence, the two UAVs hover at the altitude at which solar energy harvesting is the highest. For the RLWS agent with $\eta_1=0,\eta_2=10$, UAV $2$ ascends to $1500$m where the most solar energy can be harvested to maintain energy sustainability, while UAV $1$ descends down to $500$m to provision wireless service, indifferent to energy sustainability, which leads to its battery  depletion by the $19$-th minute as can be seen from Figure \ref{fig:2}(b). This is expected as the reward signal for training this agent does not penalize energy costs of UAV $1$. In contrast, the policy learned by the proposed CDRL agent strikes a balance: it ensures energy sustainability while slightly sacrificing the network capacity performance.  

\begin{figure}
    \centering
    \includegraphics[width=0.95\linewidth]{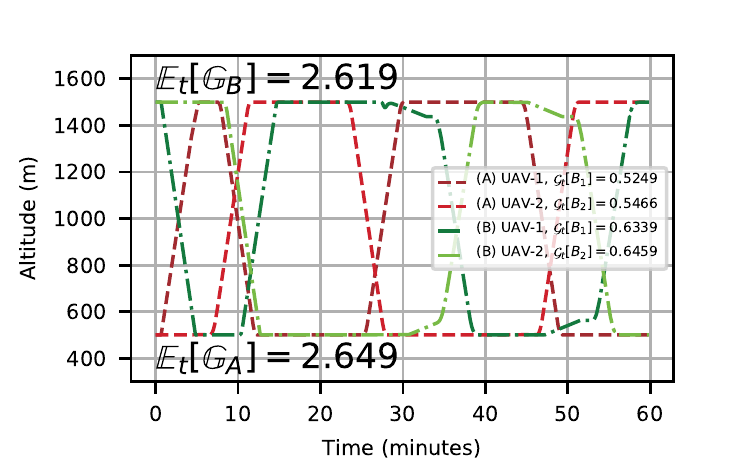}  \\  \scriptsize{(a) Policy Performance with Varying Initial Altitudes} \\ 
    \includegraphics[width=.95\linewidth]{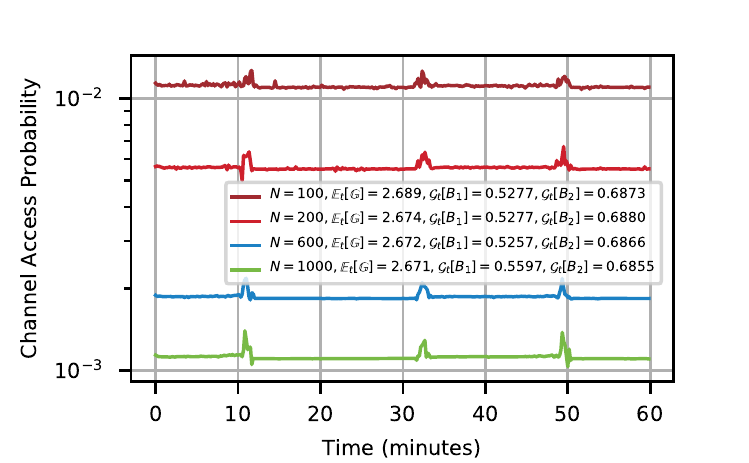} \\ \scriptsize{(b) Policy Performance with Varying Number of IoT Devices} \\
    \includegraphics[width=0.95\linewidth]{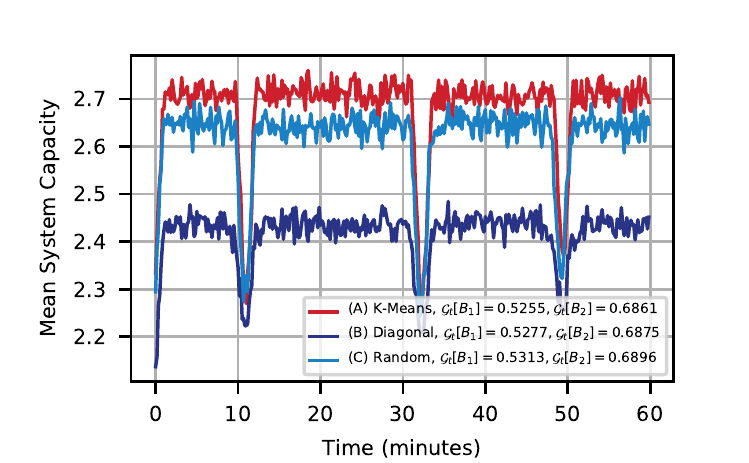}  \\  \scriptsize{(C) Policy Performance with Different  $(x^m,y^m)$ Selection Schemes} \\ 
    \caption{Learned policy generalizability with varying initial states and network scale} \label{fig:6}
  \end{figure}

\textbf{\textit{Policy Generalizability}}: To demonstrate the generalizability and robustness of the learned CDRL policy, we test its performance on networks with different initial states and varying number of IoT devices. Note that the CDRL policy has been trained given that $200$ IoT devices are uniformly deployed on ground, and that the two UAVs are initially present at altitudes of $250$m and $750$m. In Figure \ref{fig:6}(a), we demonstrate the learned policy performance given different initial altitudes of UAVs. Specifically, we consider two extreme cases: both UAVs are initially deployed at the altitude of $500$m (case A), or at $1500$m (case B). For those two cases, Figure \ref{fig:6}(a) shows that the two UAVs tend to fully de-synchronize their vertical flight trajectories, such that when one of them is charging its battery at $1500$m, the other is provisioning wireless service at $500$m. The temporal average system capacity for cases A and B are $2.649*\mathcal{W}$ bit per second (bps) and $2.619*\mathcal{W}$ bps, respectively. Notice that case B achieves a slightly lower temporal average network capacity because both UAVs are initially farther away from IoT devices. In the legend, the temporal geometric mean of the battery energy of each UAV, defined as $\mathcal{G}_t[B_m]=\big( \prod_{n=1}^H B_m(t_n) \big)^{\frac{1}{H}}$, is reported to demonstrate energy sustainability of UAVs throughout the operating horizon.   
 
In Figure \ref{fig:6}(b), we test how the learned CDRL policy scales with varying number of IoT devices. As can be seen from Figure \ref{fig:6}(b), the channel access probability is scaled appropriately given the number of IoT devices vary in $\{100,200,600,1000\}$, maintaining comparable temporal average network capacity around $2.67*\mathcal{W}$ bps. In addition, the temporal geometric mean of the battery energy of each UAV is also reported in the legend to demonstrate energy sustainability of UAVs. Last but not least, in Figure \ref{fig:6}(c), we test the performance of the learned policy given different horizontal deployment of the two UAVs. We consider three cases: (A) the two UAVs are  deployed at $(250,250)$m and $(750,250)$m, as determined by the K-means clustering algorithm, (B) the two UAVs are deployed farthest from each other at $(0,0)$m and $(1000,500)$m, and (C) the two UAVs are randomly deployed on the xy-plane. It can be observed from Figure \ref{fig:6}(c) that the mean network capacity is highest when the K-means algorithm is employed to determine the xy-planar deployment of the two UAVs. In addition, it is shown that randomly deploying the two UAVs in the xy-plane, case (C), is slightly lower than that in case (A), whereas the extreme case of deploying the two UAVs on the diagonal, case (B), achieves the lowest network capacity. In all cases, the learned policy still ensures energy sustainability of the two UAVs as indicated by the temporal geometric mean of the battery energy of each UAV, which is reported in the legend. 
 
\section{Conclusion}
In this paper, we have applied constrained deep reinforcement learning to study the joint problem of dynamic multi-UAV altitude control and random channel access management of a multi-cell UAV-based wireless network with NOMA, in support of a massive number of IoT devices. To enable an energy-sustainable capacity-optimal IoT network, we have formulated this constrained stochastic control problem as a constrained markov decision process, and proposed an online model-free constrained deep reinforcement learning algorithm to learn an  optimal control policy for wireless network management. Our extensive simulations have  demonstrated that the proposed algorithm learns a cooperative policy in which the altitude of UAVs and channel access  probability of IoT  devices are dynamically adapted to  maximize the long-term total network capacity while maintaining energy sustainability of UAVs. The policy learned by the proposed algorithm ensures energy sustainable operation of UAVs, and outperforms baseline solutions. In our future work, we will study the design of a constrained multi-agent RL framework to tackle resource management problems in spatially distributed massive wireless networks.

\section*{Acknowledgment}
This work was supported in part by the NSF grants ECCS1554576, ECCS-1610874, and CNS-1816908. We gratefully acknowledge the computing resources provided on Bebop, a high-performance computing cluster operated by the Laboratory Computing Resource Center at Argonne National Laboratory.
\pagenumbering{arabic}
\renewcommand{\thepage} {\arabic{page}}
\bibliographystyle{IEEEtran}
\bibliography{M335}
\input{bio.tex}
\end{document}

%% file: bio.tex
\begin{IEEEbiography}[{\includegraphics[width=1in,height=1.25in,clip,keepaspectratio]{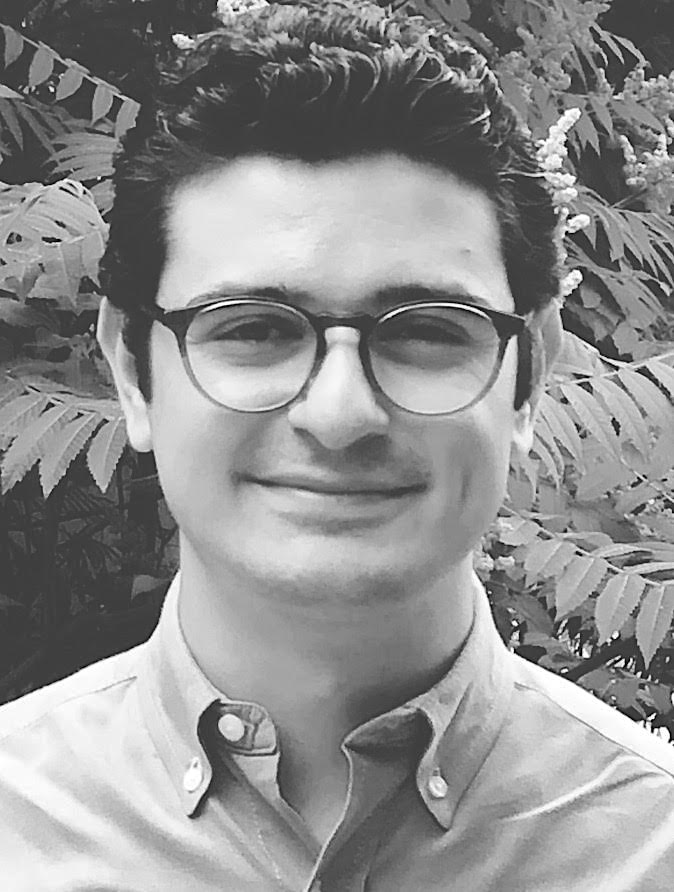}}]%
{Sami Khairy} (S’16) received the B.S. degree in Computer Engineering from the University of Jordan, Amman, Jordan, in 2014 and the M.S. degree in Electrical Engineering from Illinois Institute of Technology, Chicago, IL, USA, in 2016. He is currently working towards the Ph.D. degree in Electrical Engineering at Illinois Institute of Technology. His research interests span the broad areas of analysis and protocol design for next generation wireless networks, AI powered wireless networks resource management, reinforcement learning, statistical learning, and statistical signal processing. He received a Fulbright Predoctoral Scholarship from JACEE and the U.S. Department of State in 2015, and the Starr/Fieldhouse Research Fellowship from IIT in 2019. He is an IEEE student member and a member of IEEE ComSoc and IEEE HKN.
\end{IEEEbiography}

\begin{IEEEbiography}[{\includegraphics[width=1in,height=1.25in,clip,keepaspectratio]{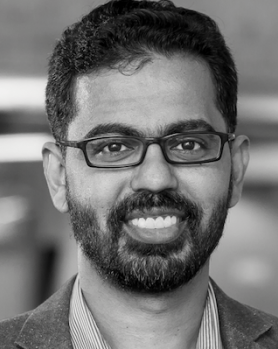}}]%
{Prasanna Balaprakash} is a computer scientist at the Mathematics and Computer Science Division with a joint appointment in the Leadership Computing Facility at Argonne National Laboratory. His research interests span the areas of artificial intelligence, machine learning, optimization, and high-performance computing. Currently, his research focuses on the development of scalable, data-efficient machine learning methods for scientific applications. He is a recipient of U.S. Department of Energy 2018 Early Career Award. He is the machine-learning team lead and data-understanding team co-lead in RAPIDS, the SciDAC Computer Science institute. Prior to Argonne, he worked as a Chief Technology Officer at Mentis Sprl, a machine learning startup in Brussels, Belgium. He received his PhD from CoDE-IRIDIA (AI Lab), Université Libre de Bruxelles, Brussels, Belgium, where he was a recipient of Marie Curie and F.R.S-FNRS Aspirant fellowships.
\end{IEEEbiography}

\begin{IEEEbiography}
[{\includegraphics[width=1in,height=1.25in,clip,keepaspectratio]{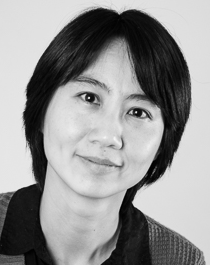}}]
{Lin X. Cai} (S’09–M’11–SM’19) received the M.A.Sc. and Ph.D. degrees in Electrical and Computer Engineering from the University of Waterloo, Waterloo, Canada, in 2005 and 2010, respectively. She is currently an Associate Professor with the Department of Electrical and Computer Engineering, Illinois Institute of Technology, Chicago, Illinois, USA. Her research interests include green communication and networking, intelligent radio resource management, and wireless Internet of Things. She received a Postdoctoral Fellowship Award from the Natural Sciences and Engineering Research Council of Canada (NSERC) in 2010, a Best Paper Award from the IEEE Globecom 2011, an NSF Career Award in 2016, and the IIT Sigma Xi Research Award in the Junior Faculty Division in 2019. She is an Associated Editor of IEEE Transaction on Wireless Communications, IEEE Network Magazine, and a co-chair for IEEE conferences.  
\end{IEEEbiography}

\begin{IEEEbiography}
[{\includegraphics[width=1in,height=1.25in,clip,keepaspectratio]{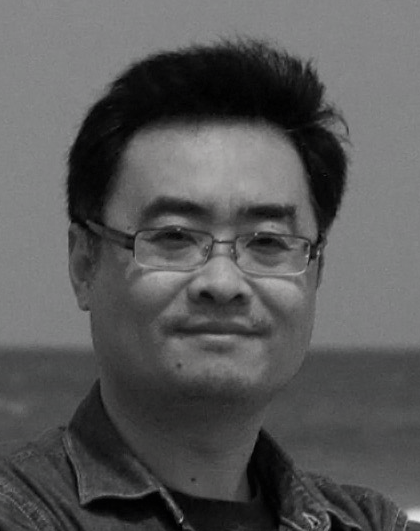}}]
{Yu Cheng} (S’01–M’04–SM’09) received the B.E. and M.E. degrees in electronic engineering from Tsinghua University, Beijing, China, in 1995 and 1998, respectively, and the Ph.D. degree in electrical and computer engineering from the University of Waterloo, Waterloo, ON, Canada, in 2003. He is currently a Full Professor with the Department of Electrical and Computer Engineering, Illinois Institute of Technology, Chicago, IL, USA. His current research interests include wireless network performance analysis, network security, big data, cloud computing, and machine learning. Dr. Cheng was a recipient of the Best Paper Award at QShine 2007, the IEEE ICC 2011, the Runner-Up Best Paper Award at ACM MobiHoc 2014, the National Science Foundation CAREER Award in 2011, and the IIT Sigma Xi Research Award in the Junior Faculty Division in 2013. He has served as several Symposium Co-Chairs for IEEE ICC and IEEE GLOBECOM, and the Technical Program Committee Co-Chair for WASA 2011 and ICNC 2015. He was a founding Vice Chair of the IEEE ComSoc Technical Subcommittee on Green Communications and Computing. He was an IEEE ComSoc Distinguished Lecturer from 2016 to 2017. He is an Associate Editor for the IEEE Transactions on Vehicular Technology, IEEE Internet of Things Journal, and IEEE Wireless Communications.
\end{IEEEbiography}